\documentstyle[12pt,epsf]{article}
\makeatletter
\@addtoreset{equation}{section}

\makeatother

\def\sp{\phantom{a}}
\def\Sp{\phantom{A}}
\def\bmi#1{\mbox{\boldmath $#1$}}
\def\Rn#1{\uppercase\expandafter{\romannumeral#1}}

\begin{document}
\sloppy
\sloppy
\sloppy

\begin{flushright}{UT-864\\November, 1999}

\end{flushright}
\vskip 0.5 truecm

\vskip 0.5 truecm

\begin{center}
{\large{\bf  Entanglement entropy of the black hole horizon}}
\end{center}
\vskip .5 truecm
\centerline{\bf Hiroaki Terashima}
\vskip .4 truecm
\centerline {\it Department of Physics,University of Tokyo}
\centerline {\it Bunkyo-ku,Tokyo 113-0033,Japan}
\vskip 0.5 truecm

\begin{abstract}
We examine a possibility that,
when a black hole is formed,
the information on the collapsed star is stored
as the entanglement entropy between 
the outside and the thin region
(of the order of the Planck length)
of the inside the horizon.
For this reason, we call this as
the entanglement entropy of the black hole ``horizon''.
We construct two models,
one is in the Minkowski spacetime and
the other is in the Rindler wedge.
To calculate the entropy explicitly,
we assume that the thin regions of the order of
the Planck length of the outside and inside the horizon
are completely entangled by quantum effects.
We also use a property of the entanglement entropy
that it is symmetric under an interchange of
the observed and unobserved subsystems.
Our setting and this symmetric property
substantially reduce the needed numerical calculation.
As a result of our analysis,
we can explain the Bekenstein-Hawking entropy itself
(rather than its correction by matter fields)
in the context of the entanglement entropy.
\end{abstract}

\section{Introduction}
\label{intro}
There is a well-known analogy between 
black hole physics and thermodynamics.
This fact is called the black hole thermodynamics~\cite{BaCaHa73}.
In particular, as first pointed out
by Bekenstein~\cite{Bekens73},
we can think of the area of the black hole horizon
as the entropy (up to a proportional constant)
by using the area theorem~\cite{AreaTh}
which states that the area of
the black hole horizon does not decrease.
Since the black hole emits thermal radiation of matter,
which is called the Hawking
radiation~\cite{Hawkin75,Wald75,Hawkin76},
we can decide the temperature of the black hole.
Thus, the entropy of the black hole is calculated as
\begin{equation}
  S=\frac{1}{4l^2_{pl}}A,
\label{BH}
\end{equation}
where $A$ is the area of the horizon and 
$l_{pl}=(\hbar G/c^3)^{1/2}$ is the Planck length.
This is the Bekenstein-Hawking entropy.

There have been many attempts
to understand the origin of this black hole entropy:
For example, those considerations on the basis of
the value of the Euclidean
action~\cite{GibHaw77,Hawkin79,BroYor93b,BaTeZa94,HawHun99},
the rate of the pair creation of black holes~\cite{GaGiSt94},
the Noether charge of the bifurcate Killing
horizon~\cite{Wald93,IyeWal94} or
the central charge of the Virasoro
algebra~\cite{Carlip95,Stromi98,Carlip99}.
Among past considerations, we consider the entanglement
entropy~\cite{BKLS86,Sredni93,FroNov93,CalWil94,%
KabStr94,HoLaWi94,MuSeKo97,MuSeKo98,Others}
as the most attractive candidate for the black hole entropy.
The entanglement entropy is the measure of the
information loss due to a division of the system;
this direct connection of the entropy with the information loss
is not clear in some other approaches to the black hole entropy.
If we divide the system into two subsystems $A$ and $B$,
and ignore the information about $B$ and observe only $A$,
we can view the pure state of the total system as
an effective mixed state for the subsystem $A$.
The entanglement entropy is the von Neumann entropy
of this effective mixed state.
If the original pure state is an entangled state,
the entanglement entropy is non-zero.
On the other hand, if the original pure state is not 
an entangled state, the entanglement entropy is zero.
That is, if the original pure state is not entangled,
there is no information loss when we ignore $B$.
Note that the entangled state and the entanglement entropy is
a purely quantum mechanical notion and
there is no counterpart in classical physics.

When the concept of the entanglement entropy
is applied to the black hole,
it measures the information loss due to a spatial separation.
Most previous works on the entanglement entropy were
concentrated on the entanglement between the bulk regions outside
and inside of the black hole horizon.
In this paper, we instead discuss the entanglement
between the outside and a thin region
(of the order of the Planck length) inside of the horizon.
For this reason, we call this as
the entanglement entropy of the black hole ``horizon''.

We consider that this approach is justified physically
by the following discussion:
For simplicity,
we consider a quantum field on the extended Schwarzschild
spacetime rather than a dynamical spacetime which describes
the gravitational collapse to the black hole.
Since we want to calculate
the entropy of the black hole itself,
the quantum state of the field must be a ``vacuum''.
We thus consider the Killing vacuum,
which is defined by using the Killing time.
(Note that the Kruskal vacuum,
which is defined by using the Kruskal time,
contains the thermal radiation of the Killing particles,
namely, the Hawking radiation.
Therefore, if we chose the Kruskal vacuum,
the resultant entropy would be considered as the
entropy of the black hole and its correction by the matter field.)
Since
the Killing vacuum is expressed as the tensor product of
the states in one asymptotically flat region \Rn{1} (outside)
and the other asymptotically flat region \Rn{2} (inside),
the entanglement entropy between the inside and outside
of the horizon becomes zero.

However, if we consider the effect of the quantum gravity,
this vanishing entanglement entropy is not true any more
since we can not divide the system sharply due to
the quantum fluctuation of the horizon.
By this correction, the Killing vacuum is deformed
to some entangled states between the inside and outside
of the (classical) horizon.
The depth of the entanglement is of the order of the
Planck length because this is the effect of the quantum gravity.
Of course, it is difficult to achieve such a calculation.
To estimate this entanglement entropy,
we first assume that the thin regions 
of the order of the Planck length outside and inside the horizon
are completely entangled by quantum effects.
Namely, the main features of the states in the thin region
inside the horizon are smoothly extrapolated from
those of the outside the horizon.
The major ansatz of our calculation is that
the entanglement entropy between the thin region
inside the horizon and all the states outside the horizon
is approximated by the entanglement entropy
of the Killing vacuum between
the thin region of the order of the Planck length
{\em outside} the horizon and the rest of
the states outside the horizon.
That is, we consider that the effect of the quantum gravity
is approximated by the shift of
the (classical) division of the system
rather than the deformation of the state.
The shift is of the order of the Planck length
because this is induced by
the quantum fluctuation of the horizon.

The present ansatz is analogous to the setting
which has been considered in Ref.\cite{MuSeKo98}
in a different context.
They have considered a thin spherical shell infalling
toward a Schwarzschild black hole and
the entanglement entropy of the Killing vacuum
associated with the division by a timelike surface
which becomes the horizon {\em after} the passage of the shell,
but it is in the Schwarzschild spacetime {\em before}
the passage of the shell.
Our case is different since we analyze the
entanglement entropy generated by quantum effects
after the formation of the classical horizon.
Also, the calculation becomes much simpler
in this paper.
The key point is that the entanglement entropy is symmetric
under an interchange of the role of the subsystems A and B.
Our setting and this symmetric property
make the calculation very simple and
substantially reduce the needed numerical calculation.
Moreover, since the calculation in this paper is
based on the Bombelli-Koul-Lee-Sorkin
type calculation~\cite{BKLS86}
rather than the Srednicki type calculation~\cite{Sredni93},
we can find directly that the entanglement entropy
is proportional to the area
without plotting the entanglement entropy to the area.

There are some comments on the above setting
of the calculation in this paper:
As is well known,
Euclidean geometry plays an important role
in the Gibbons-Hawking method~\cite{GibHaw77,Hawkin79}
or some other Euclidean approaches to the black hole entropy.
Especially,
the temperature of the black hole 
can be well understood as the period of the Euclidean time
in Euclidean geometry~\cite{GibPer,ChrDuf78,TroVan79}.
Since the entropy is the conjugate variable to 
the temperature, we want to understand it
within Euclidean geometry.
However, the Euclidean black hole does not have
the ``inside'' of the horizon~\cite{Hawkin79,Wald84}.
On the other hand,
in Euclidean gravity, the horizon is the fixed point
of the Euclidean time translation,
called the bolt~\cite{GibHaw79},
and an obstruction to the foliation by the Euclidean time.
Therefore, to achieve the Hamiltonian formulation,
we want to eliminate the degrees of freedom near
the horizon~\cite{HawHor96,HawHun99}.
For this reason,
the above setting of our calculation appears to
be reasonable if we persist on the Euclidean picture.
Then, we notice that the energy, temperature and entropy could
be understood in relation to the Euclidean time translation:
That is, we regard that the energy is its charge,
the temperature is its period and
the entropy is concerned with its fixed point.

Moreover, to reach the horizon, we need an infinite time 
if the ``time'' is measured by
the asymptotically Minkowski time (not the proper time).
Thus, we can consider that the horizon is a ``boundary'',
at least, for the observer at infinity.
However, we do not impose any boundary condition at the horizon
since we do not make any measurement there.
Consequently, it is natural to take the summation
over the state at the horizon~\cite{Hawkin76}.

The plan of this paper is as follows.
In Sec.\ref{formula}, we briefly review
the notion and basic properties of the entanglement entropy,
and then derive a basic formula to calculate it.
In Sec.\ref{model}, we construct two models and
calculate the entanglement entropy explicitly.
In Sec.\ref{conclude}, we conclude and discuss
the results of this paper.

\section{Entanglement Entropy}
\label{formula}
We review the notion and properties
of the entanglement entropy
and then derive a basic
formula~\cite{BKLS86} to calculate it.

\subsection{Definition}
Let us consider the case where the total system 
can be divided into two subsystems.
Then, the Hilbert space of the total system ${\cal H}$
can be written by the tensor product,
\begin{equation}
{\cal H}={\cal H}_1\otimes{\cal H}_2.
\end{equation}
A state $|\Psi\rangle\in{\cal H}$ is called entangled
if the state can {\em not} be written as
\begin{equation}
 |\Psi\rangle=|\psi_1\rangle|\psi_2\rangle,
\end{equation}
where $|\psi_1\rangle\in{\cal H}_1$ and
$|\psi_2\rangle\in{\cal H}_2$.
For example,
if $|a\rangle,|b\rangle\in{\cal H}_1$ and
$|\alpha\rangle,|\beta\rangle\in{\cal H}_2$,
\begin{equation}
 |\Psi\rangle=|a\rangle|\alpha\rangle
    +|b\rangle|\beta\rangle
\end{equation}
is an entangled state and
\begin{eqnarray}
 |\Psi\rangle &=& |a\rangle|\alpha\rangle
    +2|a\rangle|\beta\rangle+|b\rangle|\alpha\rangle
   +2|b\rangle|\beta\rangle \nonumber \\
    &=& \Bigl(|a\rangle+|b\rangle\Bigr)
       \Bigl(|\alpha\rangle+2|\beta\rangle\Bigr)
\end{eqnarray}
is not an entangled state.

Moreover, we assume that we are going to ignore the degrees
of freedom of ${\cal H}_2$.
To achieve this, we define a reduced density matrix
$\rho_{{\rm red}}$ for ${\cal H}_1$ from the (pure) state
of the total system $|\Psi\rangle$,
whose matrix elements are given by
\begin{equation}
  \langle a|\rho_{{\rm red}}|b \rangle =
    \sum_\alpha\, \Bigl(\langle a|\langle\alpha |\Bigr)
   |\Psi\rangle\langle\Psi|
  \Bigl(|\, b\rangle |\alpha\rangle\Bigr),
\label{reduced}
\end{equation}
where $|a \rangle,|\,b \rangle$ are the arbitrary
states of ${\cal H}_1$ and $\{|\alpha \rangle\}$
are the orthonormal basis of ${\cal H}_2$.
Then, the expectation value of an operator $O$ which acts
only on ${\cal H}_1$ becomes
\begin{equation}
  \langle \Psi | O | \Psi \rangle
  ={\rm Tr}_1\,(\rho_{{\rm red}} O),
\end{equation}
where the trace is taken over the states of ${\cal H}_1$.
By this way, as far as the subsystem ${\cal H}_1$ is concerned,
the pure state of the total system $|\Psi\rangle$ can be viewed
as the mixed state $\rho_{{\rm red}}$.

Now, the entanglement entropy is defined by the von Neumann
entropy of this reduced density matrix,
\begin{eqnarray}
   S_{12} &=&-{\rm Tr}_1 (\rho_{{\rm red}}\ln\rho_{{\rm red}})
             \nonumber \\
         &=&-\sum_n p_n\ln p_n,
\end{eqnarray}
where $\{p_n\}$ are the eigenvalues of $\rho_{{\rm red}}$.
Note that the range of the entanglement entropy is
\begin{equation}
   0\le S_{12} \le \ln N,
\end{equation}
where $N$ is the dimension of ${\cal H}_1$.

If the original state $|\Psi\rangle$ is not entangled,
$\rho_{{\rm red}}$ remains pure and,
thus, $S_{12}$ becomes zero.
On the other hand, if $|\Psi\rangle$ is entangled,
$\rho_{{\rm red}}$ becomes a mixed state and
$S_{12}$ is nonzero.
Thus, the entanglement entropy is a measure of
the entangled nature (or EPR correlation)
of the original state.

For example, let us consider the system
which consists of two spin-1/2 particles:

If a state of the system
is an EPR state,
\begin{equation}
  |\psi\rangle=\frac{1}{\sqrt{2}}
     \left(|\uparrow_1\rangle|\uparrow_2\rangle
   +|\downarrow_1\rangle|\downarrow_2\rangle\right),
\end{equation}
then the reduced density matrix becomes
\begin{equation}
\rho_{{\rm red}}=\pmatrix{ \frac{1}{2} &      0      \cr
                      0      & \frac{1}{2} \cr },
\end{equation}
and the entanglement entropy is $S_{12}=\ln2$.
This state has the maximum entanglement entropy~\cite{Mukohy98}.
Note that, since there is a perfect EPR correlation
between these particles,
we can get full information about one particle by 
an observation of the other particle.
Thus, these particles are maximally entangled.

On the other hand,
if a state of the system is not an entangled state,
\begin{eqnarray}
  |\psi\rangle &=&\frac{1}{\sqrt{2}}\left(|\uparrow_1\rangle
   +|\downarrow_1\rangle\right)
  \otimes\frac{1}{\sqrt{2}}\left(|\uparrow_2\rangle
   +|\downarrow_2\rangle\right) \nonumber \\
    &=& \frac{1}{2} 
    \left(|\uparrow_1\rangle|\uparrow_2\rangle
         +|\uparrow_1\rangle|\downarrow_2\rangle
         +|\downarrow_1\rangle|\uparrow_2\rangle
         +|\downarrow_1\rangle|\downarrow_2\rangle \right),
\end{eqnarray}
then the reduced density matrix becomes
\begin{equation}
\rho_{{\rm red}}=\pmatrix{ \frac{1}{2} & \frac{1}{2} \cr
                 \frac{1}{2} & \frac{1}{2} \cr },
\end{equation}
and the entanglement entropy is $S_{12}=0$.
This state does not have the entanglement entropy.
Note that, since there is no EPR correlation
between these particles,
we can not get any information about one particle by 
an observation of the other particle.
Thus, these particles are not entangled.

Note that one of the important properties
of the entanglement entropy is that it is symmetric
under an interchange of the role of
${\cal H}_1$ and ${\cal H}_2$, 
\begin{equation}
S_{12}=S_{21}.
\label{symm}
\end{equation}
This is because the entanglement entropy measures
the EPR ``correlation'' between two subsystems,
which is symmetric by definition.
As for more detailed analysis,
see Refs.~\cite{Sredni93,MuSeKo97}.

\subsection{Basic Formula}
Let us consider a system
which consists of coupled oscillators, $\{q^A\}$.
Now, we will calculate the entanglement entropy
of the ground state when the system is divided
into two subsystems, $\{q^a\}$ and $\{q^\alpha\}$~\cite{BKLS86}.

The Lagrangian of the total system is given by
\begin{equation}
L=\frac{1}{2} G_{AB}\,\dot{q}^A\dot{q}^B-
   \frac{1}{2} V_{AB}\,q^Aq^B.
\label{Lagra}
\end{equation}
(We assume that $G_{AB}$ and $V_{AB}$ are symmetric and
positive definite matrices of constants.)
The canonical momentum conjugate to $q^A$ is
\begin{equation}
 p_A=G_{AB}\,\dot{q}^B.
\end{equation}
By using $(G^{-1})^{AB}$ which is the inverse matrix of $G_{AB}$
defined by
\begin{equation}
  (G^{-1})^{AB}G_{BC}=\delta^A_{\Sp C},
\end{equation}
the Hamiltonian becomes
\begin{equation}
  H=\frac{1}{2}(G^{-1})^{AB} \,p_Ap_B+
   \frac{1}{2} V_{AB}\,q^Aq^B.
\end{equation}
Moreover, we define $W_{AB}$ by
\begin{equation}
(G^{-1})^{AB} W_{AC}W_{BD}=V_{CD}.
\end{equation}
That is, $W_{AB}$ is the square root of $V_{AB}$
in terms of the metric $(G^{-1})^{AB}$.
Then, by using the canonical commutation relation
\begin{equation}
  \left[ q^A,p_B \right] =i \delta^A_{\Sp B},
\end{equation}
one finds that the Hamiltonian becomes
\begin{equation}
H=\frac{1}{2}(G^{-1})^{AB}
  \left(p_A+iW_{AC}q^C\right)\left(p_B-iW_{BD}q^D\right)
   +\frac{1}{2}(G^{-1})^{AB}W_{AB}.
\end{equation}
Thus, we can define the creation operator $a^\dagger_A$ and
the annihilation operator $a_A$ by
\begin{eqnarray}
   a_A &=& \frac{1}{\sqrt{2}}\left(p_A-iW_{AC}q^C\right), \\
   a^\dagger_A &=& \frac{1}{\sqrt{2}}
                \left(p_A+iW_{AC}q^C\right).
\end{eqnarray}
The commutation relation between these operators are
\begin{equation}
  \left[ a_A, a^\dagger_B\right]=W_{AB}.
\end{equation}
We can then write the Hamiltonian as
\begin{equation}
  H=(G^{-1})^{AB}\left[\,a^\dagger_Aa_B+
      \frac{1}{2}W_{AB}\right].
\end{equation}
The first term is the number operator and
the second term is the zero-point energy.

The ground state of this system is given by
\begin{equation}
a_A|0\rangle=
\frac{1}{\sqrt{2}}\left(p_A-iW_{AC}q^C\right)|0\rangle=0.
\end{equation}
The wave function of the ground state is
obtained by
\begin{equation}
\left(\frac{\partial}{\partial q^A}+W_{AC}q^C\right)
   \langle\{q^A\}|0\rangle=0,
\end{equation}
since $p_A=-i\partial/\partial q^A$ in the
Schr\"odinger representation.
The normalized solution of this equation is
\begin{equation}
\langle\{q^A\}|0\rangle=\left[{\rm det}\frac{W}{\pi}\right]^{1/4}
  \exp\left[-\frac{1}{2}W_{AB}q^Aq^B\right].
\end{equation}
The density matrix of this ground state is
\begin{eqnarray}
 \rho\left(\{q^A\},\{q'^B\}\right)
  &=& \langle\{q^A\}|0\rangle \langle0|\{q'^B\}\rangle\\
  &=& \left[{\rm det}\frac{W}{\pi}\right]^{1/2}
       \exp\left[-\frac{1}{2}W_{AB}
        \left(q^Aq^B+q'^Aq'^B\right)\right].
\end{eqnarray}

Now, we divide the system $\{q^A\}$
into two subsystems, $\{q^a\}$ and $\{q^\alpha\}$.
If we want to ignore the information on $\{q^\alpha\}$,
we take the trace over $\{q^\alpha\}$ and consider
the reduced density matrix as Eq.(\ref{reduced}),
\begin{eqnarray}
 \rho_{{\rm red}}\left(\{q^a\},\{q'^b\}\right)
  &=& \int \prod_\alpha dq^\alpha
        \;\rho\left(\{q^a,q^\alpha\},\{q'^b,q^\alpha\}\right)
\end{eqnarray}
By dividing $W_{AB}$ into four blocks
\begin{equation}
W_{AB}=\left(\begin{array}{cc}
 A_{ab} & B_{a\beta} \\
 (B^T)_{\alpha b} & D_{\alpha\beta}
\end{array}\right),
\end{equation}
we find that
\begin{eqnarray}
 \rho_{{\rm red}}\left(\{q^a\},\{q'^b\}\right)
  &=& \left[{\rm det}\frac{M}{\pi}\right]^{1/2}
        \exp\left[-\frac{1}{2}M_{ab}
       \left(q^aq^b+q'^aq'^b\right) \right] \nonumber \\
  & &  {}\times \exp\left[-\frac{1}{4}N_{ab}
        \left(q^a-q'^a\right)\left(q^b-q'^b\right)\right],
\end{eqnarray}
where
\begin{eqnarray}
  M_{ab} &=& A_{ab}-\left(BD^{-1}B^T\right)_{ab}, \\
  N_{ab} &=& \left(BD^{-1}B^T\right)_{ab},
\end{eqnarray}
and we have used that
\begin{eqnarray*}
 {\rm det}W &=& {\rm det}\left(\begin{array}{cc}
 A_{ab} & B_{a\beta} \\
 (B^T)_{\alpha b} & D_{\alpha\beta}
\end{array}\right)   \\
  &=& {\rm det}\left(\begin{array}{cc}
 A_{ab}-\left(BD^{-1}B^T\right)_{ab} & B_{a\beta} \\
 0 & D_{\alpha\beta}
\end{array}\right)  \\
 &=& {\rm det}M{\rm det}D.
\end{eqnarray*}

Moreover, we can choose a basis $\{\tilde{q}^a\}$
in which both $M_{ab}$ and $N_{ab}$ are diagonal.
Then, in this basis, the reduced density matrix
becomes
\begin{equation}
 \rho_{{\rm red}}\left(\{\tilde{q}^a\},\{\tilde{q}'^b\}\right)
  = \prod_a\Biggl\{ \frac{1}{\sqrt{\pi}}
        \exp\left[-\frac{1}{2}
       \left(\left(\tilde{q}^a\right)^2+\left(\tilde{q}'^a\right)^2
        \right)-\frac{1}{4}\lambda_a
        \left(\tilde{q}^a-\tilde{q}'^a\right)^2
        \right]\Biggr\},
\end{equation}
where $\{\lambda_a\}$ are the eigenvalues of the operator
\begin{equation}
  \Lambda^a_{\sp b}=(M^{-1})^{ac}N_{cb}.
\label{defL1}
\end{equation}

In order to obtain a simpler expression for $\Lambda^a_{\sp b}$,
we divide the inverse matrix of total $W_{AB}$ into four blocks,
\begin{equation}
(W^{-1})^{AB}=\left(\begin{array}{cc}
 \tilde{A}^{ab} & \tilde{B}^{a\beta} \\
 (\tilde{B}^T)^{\alpha b} & \tilde{D}^{\alpha\beta}
\end{array}\right).
\end{equation}
By definition,
\begin{eqnarray}
   \tilde{A}^{ab}A_{bc}+\tilde{B}^{a\beta}(B^T)_{\beta c}
       &=& \delta^a_{\sp c}, \label{inverse1} \\
   \tilde{A}^{ab}B_{b\gamma}+\tilde{B}^{a\beta}
   D_{\beta\gamma} &=& 0. \label{inverse2}
\end{eqnarray}
From Eq.(\ref{inverse2}), we can find
\begin{equation}
 \tilde{B}^{a\beta}=-\tilde{A}^{ab}B_{b\gamma}
        (D^{-1})^{\gamma\beta}.
 \label{Btil}
\end{equation}
Combining this with Eq.(\ref{inverse1}),
we obtain that
\begin{equation}
  (M^{-1})^{ab}= \tilde{A}^{ab}.
\label{Minv}
\end{equation}
Then, from Eq.(\ref{Btil}) and Eq.(\ref{Minv}),
it is easy to see that Eq.(\ref{defL1}) becomes
\begin{equation}
 \Lambda^a_{\sp b}=-\tilde{B}^{a\beta}(B^T)_{\beta b}
    =\tilde{A}^{ac}A_{cb}-\delta^a_{\sp b}.
\label{defL2}
\end{equation}

Note that the total reduced density matrix can be 
written by the tensor product,
\begin{equation}
  \rho_{{\rm red}}=\bigotimes_a \rho_0(\lambda_a),
\end{equation}
where
\begin{equation}
\rho_0(\lambda)=
\frac{1}{\sqrt{\pi}}\exp\left[-\frac{1}{2}
  \left(q^2+q'^2\right)
-\frac{1}{4}\lambda\left(q-q'\right)^2\right].
\end{equation}
Thus, the entropy is given by
the summation with respect to each $\lambda_a$,
\begin{equation}
 S=-{\rm Tr}\rho_{{\rm red}}\ln\rho_{{\rm red}}
  =\sum_a S(\lambda_a),
\end{equation}
where
\begin{equation}
 S(\lambda)=-{\rm Tr}\rho_0(\lambda)\ln\rho_0(\lambda).
\end{equation}
To calculate $S(\lambda)$, we must obtain
the eigenvalues of $\rho_0(\lambda)$,
\begin{equation}
  \int^\infty_{-\infty}dq'\;\rho_0(\lambda;q,q')f_n(q')
   =p_nf_n(q).
\end{equation}
By using the formula for Hermite polynomials~\cite{GraRyz80},
\begin{equation}
\int^\infty_{-\infty}dx\;e^{-(x-y)^2}H_n(\alpha x)
  =\sqrt{\pi}\left(1-\alpha^2\right)^{n/2}H_n
  \left[\frac{\alpha y}{\left(1-\alpha^2\right)^{1/2}}
  \right],
\end{equation}
we find that the eigenvalues and
eigenfunctions are given by~\cite{Sredni93}
\begin{eqnarray}
  p_n    &=& (1-\mu)\mu^n,          \\
  f_n(q) &=&  \exp\left[-\frac{1}{2}\gamma q^2\right]
              H_n(\sqrt{\gamma}q),
\end{eqnarray}
where
\begin{eqnarray}
   \gamma &=& \sqrt{1+\lambda}, \\
   \mu &=& \frac{\lambda}{\left(\sqrt{1+\lambda}+1\right)^2}.
   \label{defm}
\end{eqnarray}
Then, the entropy for $\lambda$ is given by
\begin{equation}
 S(\lambda)=-\sum_np_n\ln p_n=-\ln(1-\mu)-\frac{\mu}{1-\mu}\ln\mu.
\label{defS}
\end{equation}

In summary,
in order to calculate the entanglement entropy
for the ground state of coupled oscillators,
one must first obtain the eigenvalues $\{\lambda_a\}$ of 
$\Lambda^a_{\sp b}$ in Eq.(\ref{defL2}).
Next, for each eigenvalue $\lambda_a$,
one has to calculate $\mu_a$ by Eq.(\ref{defm}).
Finally, the entanglement entropy is given by
\begin{equation}
  S=\sum_a\left[-\ln(1-\mu_a)-\frac{\mu_a}{1-\mu_a}\ln\mu_a\right].
\end{equation}

\section{Models}
\label{model}
In this section, we will construct specific models and
calculate the entanglement entropy.
We consider a free scalar field in a background spacetime.
Since the field can be viewed as a set of coupled oscillators,
we can use the formula in the previous section.

We must divide the set of oscillators into two subsets.
In most previous works,
it was divided into the oscillators outside and
inside of the black hole.
Instead, in this paper, we will divide the system into 
the oscillators completely outside and
within a thin region $\Delta$ around horizon,
based on the discussion in Sec.\ref{intro}.
Since the thin region
is induced by the quantum fluctuation of the horizon,
the width of the region $\Delta$ is of the order of
the Planck length.
Namely, $a\sim l_{pl}$, where $a$ is the width of the region.

Furthermore, we can make the calculation simpler.
If one applies the conventional calculational scheme
to our setting, we ignore the degrees of freedom
inside of $\Delta$ (near the horizon).
Instead, we here ignore the degrees of freedom
outside of $\Delta$ in this paper.
Of course, this gives the same entanglement entropy
in our setting, because the entanglement entropy is
symmetric as in Eq.(\ref{symm}).
Moreover,
since the width of the region is of the order of
the Planck length,
we can treat the field within $\Delta$ as
a single oscillator
if we assume a momentum cut-off associated with $a$,
which is of the order of the Planck scale.
(If we assume a different momentum cut-off, which is still
of the order of the Planck scale,
we must consider the field within the thin region as
a set of oscillators rather than a single oscillator.
However,
the number of the oscillators are still of the order of 1.
Although this would change the numerical value
of the coefficient of the entanglement entropy,
it would be still the same order.
Thus, the final conclusion is unaffected.)
This makes the calculation quite simple.
Especially, the matrix $\Lambda^a_{\sp b}$ becomes
$1\times1$ matrix and the eigenvalue $\lambda_a$, itself.

\subsection{Simple Model}
\begin{figure}
\begin{center}
\leavevmode
\epsfysize=7cm
\epsfbox{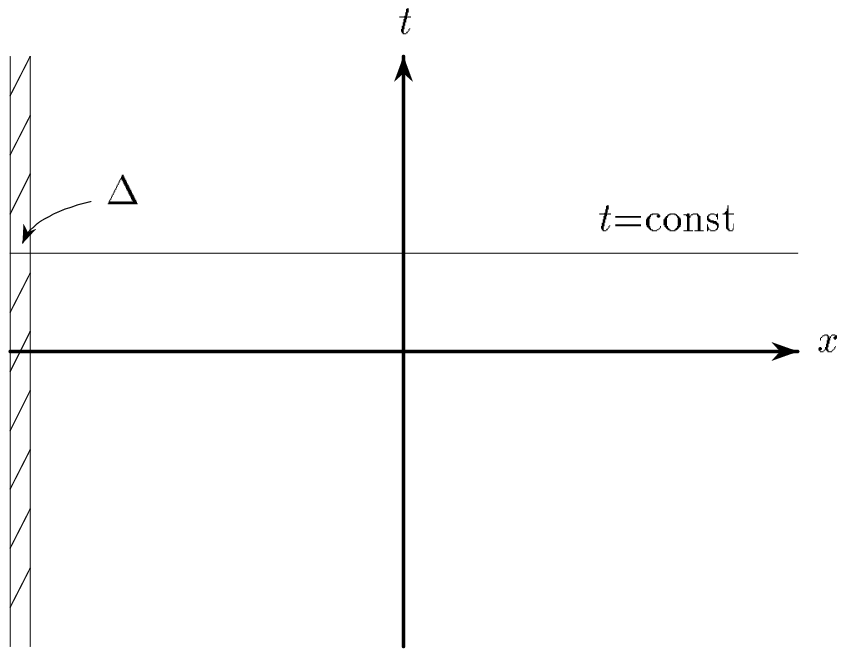}
\caption{Simple Model}
\label{model1}
\end{center}
\end{figure}

First, we consider a free scalar field in the flat spacetime and
adopt the Minkowski coordinates $(t,x,y,z)=(t,\vec{x})$.
The index $A$ is now replaced by $\vec{x}$.
We assume that the ``horizon'' is at $x=-L$ $(\to-\infty)$.
Then, the thin region near the horizon becomes
$\Delta=\{(x,y,z) | -L\le x \le -L+a \}$.
(See Fig.\ref{model1}.)
Since the action is
\begin{equation}
  S=\int dtd^3\vec{x}\; \frac{1}{2} \,
  \left[ (\partial_t\phi)^2-
         (\vec{\partial}\phi)^2-m^2\phi^2 \right],
\end{equation}
one can easily find that $G_{AB}$ and $V_{AB}$ appearing
in Eq.(\ref{Lagra}) become~\cite{BKLS86}
\begin{eqnarray}
  G(\vec{x},\vec{x}')&=& G^{-1}(\vec{x},\vec{x}')=
                          \delta(\vec{x}-\vec{x}'), \\
  V(\vec{x},\vec{x}')&=&\int\frac{d^3\vec{k}}{(2\pi)^3}
    \left(|\vec{k}|^2+m^2\right)
   \;e^{i\vec{k}\cdot (\vec{x}-\vec{x}')}.
\end{eqnarray}
Then, one finds that
\begin{eqnarray}
  W(\vec{x},\vec{x}') &=& \int\frac{d^3\vec{k}}{(2\pi)^3}
    \left(|\vec{k}|^2+m^2\right)^{1/2}
   \;e^{i\vec{k}\cdot (\vec{x}-\vec{x}')}, \\
  W^{-1}(\vec{x},\vec{x}')&=&\int\frac{d^3\vec{k}}{(2\pi)^3}
    \left(|\vec{k}|^2+m^2\right)^{-1/2}
   \;e^{i\vec{k}\cdot (\vec{x}-\vec{x}')}.
\end{eqnarray}
Thus, $\Lambda^a_{\sp b}$ in Eq.(\ref{defL2}) becomes
\begin{equation}
  \Lambda(\vec{x},\vec{x}') = 
   \int_\Delta d\vec{x}'' \left[ W^{-1}(\vec{x},\vec{x}'')
    W(\vec{x}'',\vec{x}')  \right]
  -\delta(\vec{x}-\vec{x}'),
\end{equation}
where $\vec{x},\vec{x}'\in \Delta$.

To solve the eigenvalue equation,
\begin{equation}
  \int_\Delta d\vec{x}' \Lambda(\vec{x},\vec{x}') F(\vec{x}')
   =\lambda  F(\vec{x}),
\end{equation}
we make the ansatz
\begin{equation}
  F(\vec{x}) = e^{i\bmi{p}\cdot\bmi{x}} f(x),
\end{equation}
where $\bmi{x}=(y,z)$ and $\bmi{p}=(p_y,p_z)$.
Then, the eigenvalue equation reduces to
\begin{eqnarray}
  \int^{-L+a}_{-L} dx' \int^{-L+a}_{-L} dx''
  \int \frac{dk_x}{2\pi}\int \frac{dk'_x}{2\pi}
   \times \qquad\qquad\quad
   & & \nonumber \\
  \frac{1}{\sqrt{k_x^2+M_p^2}}\sqrt{k_x^{'2}+M_p^2}
  \;e^{ik_x(x-x'')}\;e^{ik'_x(x''-x')}\, f(x')
   &=&  (\lambda+1) f(x),
\end{eqnarray}
where
\begin{equation}
 M_p  \equiv \sqrt{m^2+|\bmi{p}|^2}.
\end{equation}

\begin{figure}
\begin{center}
\leavevmode
\epsfysize=7cm
\epsfbox{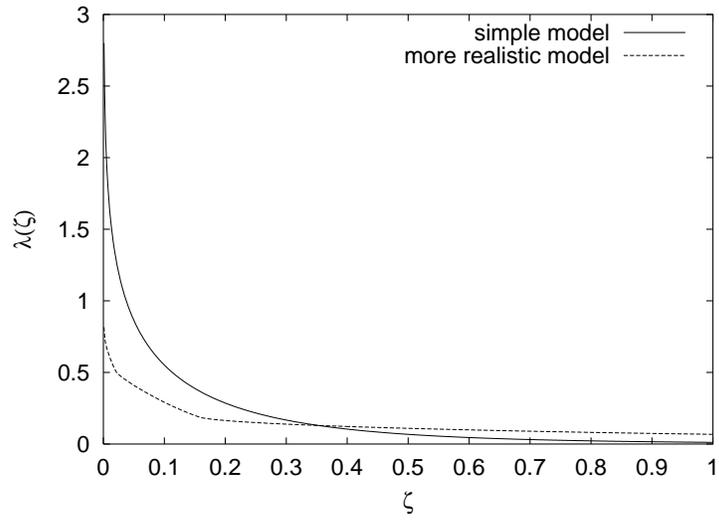}
\caption{The numerical evaluation for $\lambda(\zeta)$}
\label{fig-lam}
\end{center}
\end{figure}

\begin{figure}
\begin{center}
\leavevmode
\epsfysize=7cm
\epsfbox{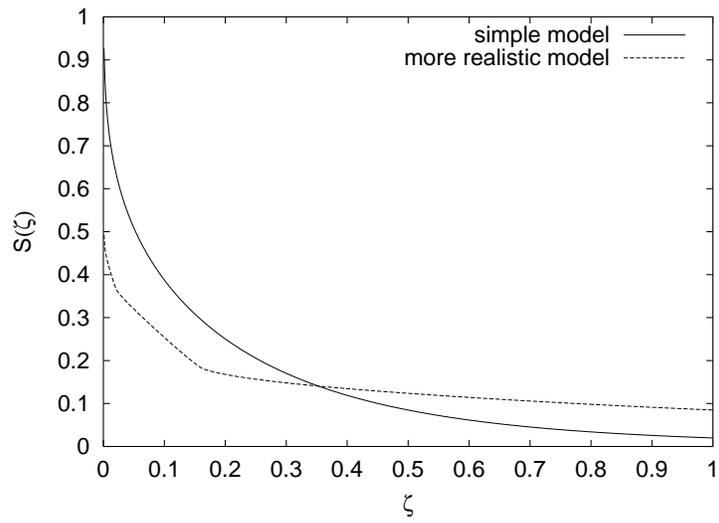}
\caption{The numerical evaluation for $S(\zeta)$}
\label{fig-ent}
\end{center}
\end{figure}

Moreover, we use an approximation,
\begin{equation}
  \int^{-L+a}_{-L} dx\, G(x) \sim a\,G(-L+a/2),
\label{approx}
\end{equation}
which corresponds to the prescription that we treat the field
within $\Delta$ as a single oscillator.
We then find that
\begin{equation}
 \lambda+1= a^2 \int \frac{dk_x}{2\pi}\int \frac{dk'_x}{2\pi}
   \frac{1}{\sqrt{k_x^2+M_p^2}}\sqrt{k_x^{'2}+M_p^2}.
\end{equation}
This would diverge unless we introduce a momentum cut-off $k_c$.
The momentum cut-off can be decided by
\begin{eqnarray}
 1&=&\int^{-L+a}_{-L}dx \int^\infty_{-\infty} \frac{dk_x}{2\pi}
   \,e^{ik_x[x-(-L+a/2)]} \nonumber \\
  &\sim & a\,\int^{k_c}_{-k_c} \frac{dk_x}{2\pi} \nonumber \\
  &=& \frac{k_c\,a}{\pi},
\label{cutoff}
\end{eqnarray}
in relation to the width of the region
and the approximation Eq.(\ref{approx}).
Thus, 
\begin{equation}
k_c=\frac{\pi}{a}.
\end{equation}
By using this momentum cut-off, we obtain that
\begin{eqnarray}
  \lambda(\zeta) &=&\frac{1}{4}
  \ln\left|\frac{\sqrt{1+\zeta^2}+1}{\sqrt{1+\zeta^2}-1}\right|
  \nonumber \\
   & & {}\times
  \left[\sqrt{1+\zeta^2}+\frac{1}{2}\zeta^2
 \ln\left|\frac{\sqrt{1+\zeta^2}+1}{\sqrt{1+\zeta^2}-1}\right|
 \right]-1,
\end{eqnarray}
where $\zeta=M_pa/\pi$.
Then, from Eq.(\ref{defm}) and Eq.(\ref{defS}),
one finds that
\begin{equation}
  \mu(\zeta) = \frac{\lambda(\zeta)}
      {\left[\sqrt{1+\lambda(\zeta)}+1\right]^2}
\end{equation}
and
\begin{equation}
  S(\zeta)=-\ln[1-\mu(\zeta)]
    -\frac{\mu(\zeta)}{1-\mu(\zeta)}\ln\mu(\zeta).
\end{equation}
Fig.\ref{fig-lam} and Fig.\ref{fig-ent}
show $\lambda(\zeta)$ and $S(\zeta)$,
respectively.

Finally, we must integrate over $\bmi{p}$ (or $\zeta$).
Note that, for a surface area $A$ in configuration space,
the density of modes in momentum space is $A/(2\pi)^2$.
(Since the shape of the ``horizon'' is $R^2$ rather than $S^2$
in this model, $A$ and the total entropy are infinite.
However, we can perfectly define the entropy per unit area,
and consequently we can pretend as if $A$ is finite
in our formula for $S$.)
Therefore,
\begin{eqnarray}
  S &=&\frac{A}{(2\pi)^2}\int d^2\bmi{p} \;S(\zeta) \nonumber \\
    &=&\frac{A}{2\pi}\int^{\pi/a}_0 p\,dp\; S(\zeta) \nonumber \\
    &\sim &\frac{\pi A}{2a^2 }\int^1_0 
         \zeta\, d\zeta \;S(\zeta),
\end{eqnarray}
where we have used the assumption that $ma\sim ml_{pl}\ll 1$.
Note that, even though $S(\zeta)\to\infty$ for $\zeta\to0$,
$\zeta S(\zeta)$ becomes zero at $\zeta=0$.
Thus, one finds
\begin{equation}
  S\sim C\,\frac{A}{a^2},
\end{equation}
where
\begin{equation}
  C=\frac{\pi}{2}\int^1_0 
    \zeta\, d\zeta \;S(\zeta)\sim 0.057.
\end{equation}
If we consider that the quantum fluctuation of the horizon is
\begin{equation}
  a \sim 2\sqrt{C}\; l_{pl} \sim 0.48 \times l_{pl} ,
\end{equation}
then the entanglement entropy is consistent with 
the Bekenstein-Hawking entropy Eq.(\ref{BH}).

\subsection{More Realistic Model}
Next, we consider a free scalar field in the flat spacetime but
adopt the Rindler coordinates
$(\tau,\xi,y,z)=(\tau,\xi,\bmi{x})$, which are defined by
\begin{eqnarray}
  t&=&\xi\sinh \alpha\tau,  \nonumber \\
  x&=&\xi\cosh \alpha\tau,
\end{eqnarray}
where $\alpha$ is a constant~\cite{Rindle66}.
The Rindler coordinates cover only
a quarter of the Minkowski spacetime, $x>|t|$,
called the Rindler wedge.
The boundary of this Rindler wedge $\xi=0$ is
the horizon for a uniformly accelerated observer
in the Rindler wedge.
In the Rindler coordinates, the flat metric becomes
\begin{equation}
  ds^2=-dt^2+dx^2+dy^2+dz^2=
   -\xi^2\alpha^2 d\tau^2+d\xi^2+dy^2+dz^2.
\label{Rindler}
\end{equation}

On the other hand,
the most general, static and spherically symmetric black hole
in four dimensions is
\begin{equation}
ds^2=-f(r)dt^2+\frac{1}{f(r)}dr^2+
\hat{R}^2(r)\,(d\theta^2+\sin^2\theta d\phi^2),
\label{4DBH}
\end{equation}
where the horizon is at $r=r_h$ which satisfies $f(r_h)=0$.
We thus make the coordinate transformation
$(t,r,\theta,\phi)\to(t,\eta,\theta,\phi)$,
which is defined by
\begin{equation}
\eta\equiv\frac{1}{\kappa}\sqrt{f},\qquad
    d\eta=\frac{1}{2\kappa}\frac{\partial_r f}{\sqrt{f}} dr,
\end{equation}
where 
\begin{equation}
\kappa=\frac{1}{2}\partial_r f|_{r=r_h}
\end{equation}
is the surface gravity of the black hole.
Note that the horizon is at $\eta=0$.
The metric (\ref{4DBH}) becomes
\begin{equation}
 ds^2=-\eta^2\kappa^2\, dt^2+\frac{4\kappa^2}{(\partial_r f)^2}d\eta^2
   +\hat{R}^2(r)\,(d\theta^2+\sin^2\theta d\phi^2),
\end{equation}
especially near the horizon $\eta\to0$,
\begin{equation}
 ds^2\to-\eta^2\kappa^2\, dt^2+d\eta^2
   +\hat{R}^2(r)\,(d\theta^2+\sin^2\theta d\phi^2).
\label{Black}
\end{equation}
Therefore, by the comparison of Eq.(\ref{Rindler})
with Eq.(\ref{Black}),
we can think of the Rindler wedge as the model for
the black hole, even though the shape of the horizon
is now $R^2$ rather than $S^2$.

\begin{figure}
\begin{center}
\leavevmode
\epsfysize=7cm
\epsfbox{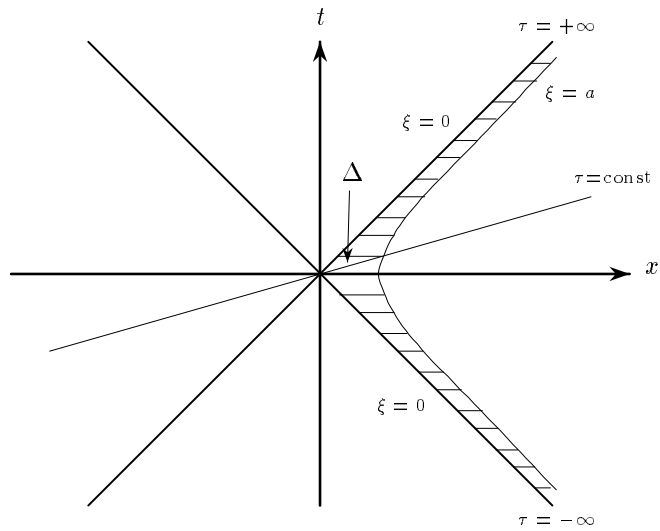}
\caption{More Realistic Model}
\label{model2}
\end{center}
\end{figure}

Since we think of the Rindler time $\tau$ as the ``time'',
the index $A$ is now replaced by $(\xi,\bmi{x})$.
The horizon is at $\xi=0$ and 
the thin region near the horizon becomes
$\Delta=\{(\xi,\bmi{x}) | 0\le \xi \le a \}$.
(See Fig.\ref{model2}.)
The ``ground state'' is the Rindler vacuum
(rather than the Minkowski vacuum)
which corresponds to the Killing vacuum
in the case of a black hole.
This is because the ``time'' is the Rindler time $\tau$
rather than the ordinary Minkowski time $t$.

Note that this horizon $\xi=0$ is a null surface,
similar to the model of Ref.~\cite{MuSeKo98}.
Thus, this model is not influenced by
the criticism~\cite{FroNov93} which is related
to the fact that the boundary of
the previous works like Refs.~\cite{BKLS86,Sredni93}
was timelike rather than null.

The action of the scalar field in the Rindler coordinates
becomes
\begin{equation}
 S = \int d\tau d\xi d^2\bmi{x}\;\frac{1}{2} 
    \xi \alpha\,\left[ \frac{1}{\xi^2\alpha^2}
   (\partial_\eta\phi)^2-(\partial_\xi\phi)^2
  -(\bmi{\partial}\phi)^2-m^2\phi^2 \right].
\end{equation}
Then, by using the orthogonality relations~\cite{Erdely54},
\begin{eqnarray}
 \frac{1}{\pi^2}\int_0^{\infty}\frac{dx}{x}\;
   K_{i\mu}(x)K_{i\nu}(x)
   &=& \frac{\delta(\mu-\nu)}{2\nu\sinh\pi\nu}, \\
 \frac{1}{\pi^2}\int_0^{\infty}d\nu\;(2\nu\sinh\pi\nu)\,
   K_{i\nu}(x)K_{i\nu}(x')
   &=& x\;\delta(x-x'),
\label{ortho}
\end{eqnarray}
which are used in the Rindler quantization~\cite{Fullin73},
one finds that $G_{AB}$ and $V_{AB}$ appearing
in Eq.(\ref{Lagra}) become
\begin{eqnarray}
  G(\xi,\bmi{x};\xi',\bmi{x}')&=& \frac{1}{\xi\alpha}
     \,\delta(\xi-\xi')\,\delta(\bmi{x}-\bmi{x}'), \\
  G^{-1}(\xi,\bmi{x};\xi',\bmi{x}')   &=&
   \xi\alpha\;\delta(\xi-\xi')\,\delta(\bmi{x}-\bmi{x}'), \\
  V(\xi,\bmi{x};\xi',\bmi{x}')&=& \frac{\alpha}{\xi\xi'}
    \int\frac{d\nu}{\pi^2}
     \int \frac{d^2\bmi{k}}{(2\pi)^2}
      \;(2\nu\sinh\pi\nu) \nonumber \\
    & & {}\times
    \nu^2\,K_{i\nu}(M_k\xi)\,K_{i\nu}(M_k\xi')
   \;e^{i\bmi{k}\cdot (\bmi{x}-\bmi{x}')}.
\end{eqnarray}
Then, one obtains that
\begin{eqnarray}
  W(\xi,\bmi{x};\xi',\bmi{x}') &=&\frac{1}{\xi\xi'}
    \int\frac{d\nu}{\pi^2}
     \int \frac{d^2\bmi{k}}{(2\pi)^2}
      \;(2\nu\sinh\pi\nu) \nonumber \\
    & & {}\times
    \nu \,K_{i\nu}(M_k\xi)\,K_{i\nu}(M_k\xi')
   \;e^{i\bmi{k}\cdot (\bmi{x}-\bmi{x}')}, \\
  W^{-1}(\xi,\bmi{x};\xi',\bmi{x}')&=&
   \int\frac{d\nu}{\pi^2}
     \int \frac{d^2\bmi{k}}{(2\pi)^2}
      \;(2\nu\sinh\pi\nu) \nonumber \\
    & & {}\times
      \nu^{-1} \,K_{i\nu}(M_k\xi)\,K_{i\nu}(M_k\xi')
   \;e^{i\bmi{k}\cdot (\bmi{x}-\bmi{x}')}.
\end{eqnarray}
Thus, $\Lambda^a_{\sp b}$ in Eq.(\ref{defL2}) becomes
\begin{eqnarray}
  \Lambda(\xi,\bmi{x};\xi',\bmi{x}') &=& 
   \int_\Delta d\xi'' d\bmi{x}'' 
    \left[ W^{-1}(\xi,\bmi{x};\xi'',\bmi{x}'')
    W(\xi'',\bmi{x}'';\xi',\bmi{x}')  \right]  \nonumber \\
    & & \qquad\qquad\qquad\qquad
  {}-\delta(\xi-\xi')\,\delta(\bmi{x}-\bmi{x}'),
\end{eqnarray}
where $(\xi,\bmi{x}),(\xi',\bmi{x}')\in \Delta$.

\begin{figure}
\begin{center}
\leavevmode
\epsfysize=7cm
\epsfbox{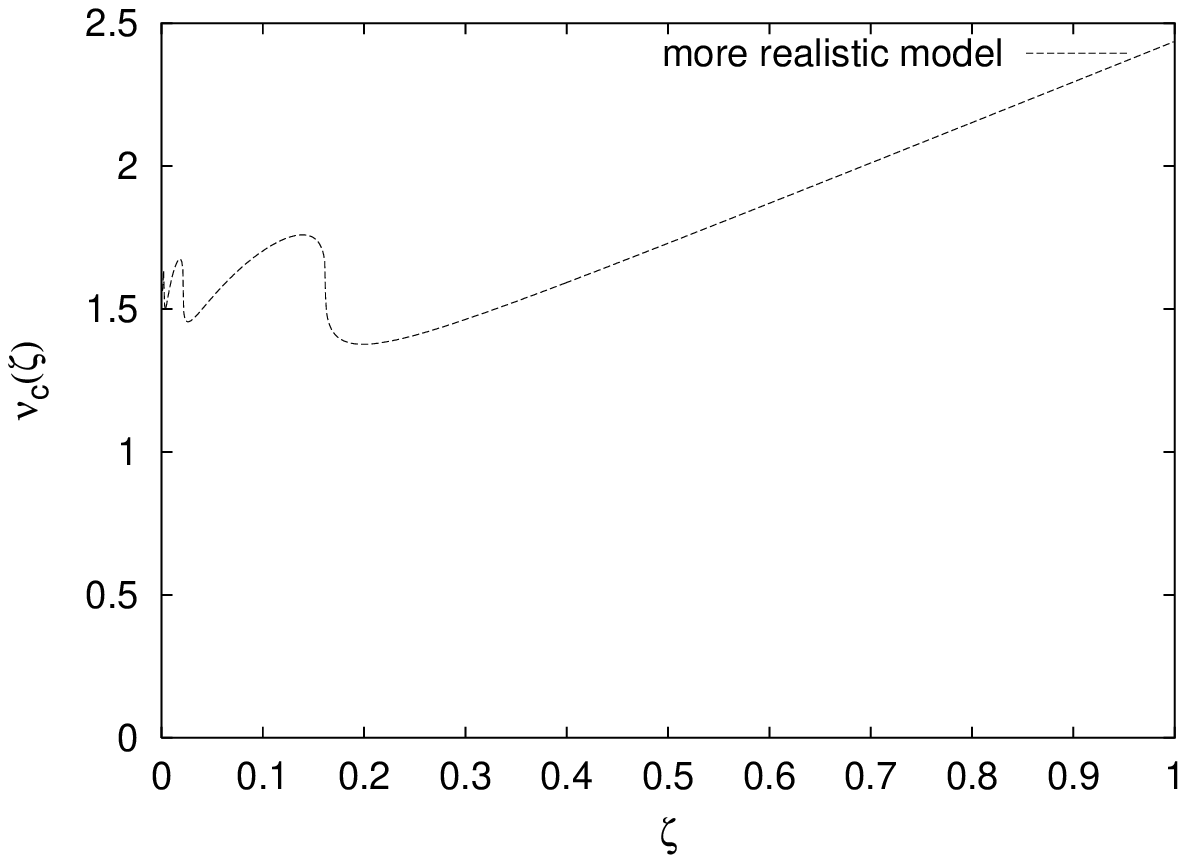}
\caption{The numerical evaluation for $\nu_c(\zeta)$}
\label{fig-cut}
\end{center}
\end{figure}

By making the ansatz as above,
\begin{equation}
  F(\xi,\bmi{x}) = e^{i\bmi{p}\cdot\bmi{x}} f(\xi),
\end{equation}
the eigenvalue equation reduces to
\begin{eqnarray}
 \int^a_0 \frac{d\xi'}{\xi'} \int^a_0 \frac{d\xi''}{\xi''}
  \int\frac{d\nu}{\pi^2}  (2\nu\sinh\pi\nu)
\int\frac{d\nu'}{\pi^2} (2\nu'\sinh\pi\nu')  \times 
 & &  \nonumber \\
\frac{\nu'}{\nu}\,
  K_{i\nu}(M_p\xi)\,K_{i\nu}(M_p\xi'')\,
   K_{i\nu'}(M_p\xi'')\,K_{i\nu'}(M_p\xi')\, f(\xi')
 & &   \nonumber \\
= &(\lambda+1)f(\xi)  & .
\end{eqnarray}
Then, by using the approximation,
\begin{equation}
  \int^a_0 d\xi\, G(\xi) \sim a\,G(a/2),
\label{approx2}
\end{equation}
we find that
\begin{eqnarray}
 \lambda+1 &=& 4\int\frac{d\nu}{\pi^2}(2\nu\sinh\pi\nu)
\int\frac{d\nu'}{\pi^2} (2\nu'\sinh\pi\nu') \nonumber \\
  &  & \qquad\qquad{}\times \frac{\nu'}{\nu}\,
\left[K_{i\nu}(M_pa/2)\right]^2
\left[K_{i\nu'}(M_pa/2)\right]^2 .
\end{eqnarray}
This would diverge unless we introduce a momentum cut-off for
$\nu$ and $\nu'$ integrals.
As in Eq.(\ref{cutoff}),
we decide the momentum cut-off in relation to
the width of the region
and the approximation Eq.(\ref{approx2}).
By using (\ref{ortho}), it can be decided by
\begin{eqnarray}
 1&=&\int^a_0 \frac{d\xi}{\xi} 
    \int^\infty_0 \frac{d\nu}{\pi^2}(2\nu\sinh\pi\nu) 
    K_{i\nu}(M_p\xi)\,K_{i\nu}(M_pa/2)\nonumber \\
  &\sim & 2\,\int^{\nu_c}_0 \frac{d\nu}{\pi^2}(2\nu\sinh\pi\nu) 
    \left[K_{i\nu}(M_pa/2)\right]^2.
\end{eqnarray}
Unfortunately, this integral can not be done analytically.
However, we can perform the numerical integration.
Note that the cut-off is not a constant but a function 
of $\zeta=M_pa/\pi$ by the dimensional analysis.
Thus,
\begin{eqnarray}
  \lambda(\zeta)&=&4\Biggl\{\int^{\nu_c(\zeta)}_0
     \frac{d\nu}{\pi^2}(2\nu\sinh\pi\nu)
   \int^{\nu_c(\zeta)}_0\frac{d\nu'}{\pi^2} (2\nu'\sinh\pi\nu') 
   \nonumber \\ 
  &  & 
  \qquad\qquad{}\times \frac{\nu'}{\nu}\,
 \left[K_{i\nu}(\pi\zeta/2)\right]^2
  \left[K_{i\nu'}(\pi\zeta/2)\right]^2\Biggr\}-1 ,
\end{eqnarray}
where $\nu_c(\zeta)$ is defined by
\begin{equation}
\int^{\nu_c(\zeta)}_0 \frac{d\nu}{\pi^2}(2\nu\sinh\pi\nu) 
    \left[K_{i\nu}(\pi\zeta/2)\right]^2=\frac{1}{2}.
\end{equation}
Then, from Eq.(\ref{defm}) and Eq.(\ref{defS}),
one obtains $\mu(\zeta)$ and $S(\zeta)$, as above.
$\lambda(\zeta)$ and $S(\zeta)$
are shown in Fig.\ref{fig-lam} and Fig.\ref{fig-ent},
respectively.
(Note that $\lambda(\zeta)$ and $S(\zeta)$
seem to be not smooth at $\zeta\sim0.02$ or $0.16$.
However, this is because the cut-off $\nu_c(\zeta)$,
which is shown in Fig.\ref{fig-cut}, varies
so rapidly there.
Thus, we need more accuracy at such points.)

Finally, after integrating over $\bmi{p}$ (or $\zeta$)
by using the fact that the density of modes in
momentum space is $A/(2\pi)^2$,
one finds that
\begin{equation}
  S\sim C\,\frac{A}{a^2},
\end{equation}
where
\begin{equation}
  C=\frac{\pi}{2}\int^1_0 
    \zeta\, d\zeta \;S(\zeta)\sim 0.089.
\end{equation}
(Even though $A$ and $S$ in our formula,
if literally taken, are infinite since
the shape of the horizon is $R^2$ rather than
$S^2$ in this model, we can still define the
entropy per unit area precisely.
We can thus pretend as if $A$ and $S$ are finite
in our final formula.)
If we consider that the quantum fluctuation of the horizon is
\begin{equation}
  a \sim 2\sqrt{C}\; l_{pl} \sim 0.60 \times l_{pl} ,
\end{equation}
then the entanglement entropy is consistent with 
the Bekenstein-Hawking entropy Eq.(\ref{BH}).

\section{Conclusion and Discussion}
\label{conclude}
In this paper, we have considered
the entanglement entropy between
the outside and the thin region
(of the order of the Planck length)
of the inside the horizon
based on the discussion in Sec.\ref{intro}.
By constructing two models, a simple one and a more realistic one,
we have shown that its entanglement entropy becomes
\begin{equation}
 S\sim C\,\frac{A}{a^2},
\end{equation}
where $a$ is the quantum fluctuation of the horizon and
$C$ is a constant.
If the quantum fluctuation of the horizon is
\begin{equation}
  a \sim 2\sqrt{C}\; l_{pl},
\end{equation}
we can interpret the Bekenstein-Hawking entropy, Eq.(\ref{BH}),
in the context of the entanglement entropy.
This is consistent with the assumption that
the quantum fluctuation of the horizon
is of the order of the Planck length.

Although some authors have considered
the entanglement entropy as the {\em correction} to
the Bekenstein-Hawking entropy generated by matter fields,
we want to consider this entanglement entropy
as the Bekenstein-Hawking entropy {\em itself}.
This is because we have considered the entanglement entropy
of the Rindler vacuum (rather than the Minkowski vacuum)
in the second model, which does not contain the thermal radiation
of the Rindler particles.
In the case of a black hole, 
this corresponds to the Killing vacuum
(rather than the Kruskal vacuum),
which does not contain the Hawking radiation.
Thus, this entropy is not associated with the existence
of the thermal radiation of particles
but rather with the existence of the black hole itself.

One might think that this entanglement entropy
would depend on the number of matter fields which are present
in the real world.
That is, if there are $N$ matter fields independently,
one might think that
the entanglement entropy would be multiplied by $N$
and conclude that this entropy
could not be considered as the entropy of
the ``black hole'', since it would depend on $N$.
However, the entanglement entropy of the horizon
in fact does {\em not} depend on $N$.
This is because the quantum fluctuation of the horizon $a$
also depends on $N$ and, besides,
it is proportional to $\sqrt{N}$.
This can be seen from the following argument.
Let us consider a Schwarzschild black hole with its mass $M$,
which fluctuates within $\delta M$ ($\delta M/M\ll 1$).
Then, the Schwarzschild radius of this black hole fluctuates
within $2\delta M$ in the {\em coordinate} length.
The {\em proper} length of this fluctuation becomes
\begin{equation}
\int^{r=2(M+\delta M)}_{r=2M}ds=\int^{r=2(M+\delta M)}_{r=2M}
\frac{dr}{\sqrt{1-2M/r}} \sim 2\sqrt{2M\,\delta M}.
\end{equation}
Note that $\delta M$ is proportional to $N$,
since the rate of spontaneous quantum emission or
absorption of particles is proportional to $N$.
Thus, the fluctuation of the horizon is proportional to
$\sqrt{N}$ in the proper length.
(This is similar to the ``brick wall''
of 't Hooft~\cite{tHooft85}.)
Since the coefficient in front of $d\xi^2$ is $1$ in
Eq.(\ref{Rindler}), $a$ is the proper length and thus
is proportional to $\sqrt{N}$.
Therefore,
if the species of matter fields becomes $N$,
the entanglement entropy of the horizon becomes
\begin{equation}
N\times
C\,\frac{A}{(\sqrt{N}a)^2}=C\,\frac{A}{a^2},
\end{equation}
which is independent of $N$~\cite{FroNov93}.
We thus consider this entanglement entropy of the horizon
as the entropy of the ``black hole'' itself rather than
the ``matter field''.

The result of our analysis suggests that
we can consider that the information on the collapsed
star is stored as the EPR correlation between
the outside and neighborhood
(of the order of the Planck length) of the horizon.
Since the horizon remains stable to the Planck scale,
we can encode the enormous information on
the collapsed star.
If we used an ordinary wall,
we could not do so because it begins to fluctuate
far below the Planck scale.
The information available outside the horizon is
the probability distribution of the effective states
(the effective density matrix)
when we ignore the field near the horizon.
Note that
this consideration does not contradict with
the no-hair theorem.

Moreover, this picture appears to be consistent with the Euclidean
path-integral approach by Gibbons and
Hawking~\cite{GibHaw77,Hawkin79,HawHun99}.
The entropy in that approach arises from the fixed point
of the Euclidean time translation or non-trivial topology
of $(\tau,r)$ section.
In our analysis, we find that the quantities
which appear in the first law
of the black hole thermodynamics
can be understood in relation to
the Euclidean time translation:
That is, the energy is its charge,
the temperature is its period and
the entropy is concerned with its fixed point.

Finally, 
to be more realistic,
we have to consider the case where the shape of
the horizon is $S^2$, like Eq.(\ref{Black}).
We then have to expand the field by the spherical harmonics
$Y_{lm}(\theta,\phi)$.
However, we expect that this would not change the result
drastically and would turn out to be
 consistent with the result of Ref.~\cite{MuSeKo98},
\begin{equation}
  S \sim 0.024 \times \frac{A}{a^2}.
\end{equation}
This is because as long as the radius of the sphere is much larger
than the Planck length $l_{pl}$
(which is equivalent to the near-horizon limit),
we can approximate the horizon as a plane.
Of course,
by using the method developed in this paper
which is based on the Bombelli-Koul-Lee-Sorkin
type calculation~\cite{BKLS86}
rather than the Srednicki type calculation~\cite{Sredni93},
we will be able to obtain the result in a much simpler
and more direct way.
This is left for a future work.

\section*{Acknowledgment}
The author thanks K. Fujikawa for a careful reading of
the manuscript and valuable comments.

%%%%%%%%%%%%%%%%%%%%%%%%%%%%%%%%%%%%%%%%%%%%%%%%%%%%%%%%%%%%%%%%%%
%   5 Figures
%%%%%%%%%%%%%%%%%%%%%%%%%%%%%%%%%%%%%%%%%%%%%%%%%%%%%%%%%%%%%%%%%%

\newpage

\begin{thebibliography}{10}

\bibitem{BaCaHa73}
J.~M. Bardeen, B. Carter, and S.~W. Hawking, Commun. Math. Phys. {\bf 31},  161
   (1973).

\bibitem{Bekens73}
J.~D. Bekenstein, Phys. Rev. {\bf D7},  2333  (1973).

\bibitem{AreaTh}
S.~W. Hawking, Phys. Rev. Lett. {\bf 26},  1344  (1971);
Commun. Math. Phys. {\bf 25},  152  (1972).

\bibitem{Hawkin75}
S.~W. Hawking, Commun. Math. Phys. {\bf 43},  199  (1975).

\bibitem{Wald75}
R.~M. Wald, Commun. Math. Phys. {\bf 45},  9  (1975).

\bibitem{Hawkin76}
S.~W. Hawking, Phys. Rev. {\bf D14},  2460  (1976).

\bibitem{GibHaw77}
G.~W. Gibbons and S.~W. Hawking, Phys. Rev. {\bf D15},  2752  (1977).

\bibitem{Hawkin79}
S.~W. Hawking,  in {\em General Relativity, an {E}instein Centenary Survey},
  edited by S.~W. Hawking and W. Israel (Cambridge University Press, Cambridge,
  1979).

\bibitem{BroYor93b}
J.~D. Brown and J.~W. York, Phys. Rev. {\bf D47},  1420  (1993), gr-qc/9209014.

\bibitem{BaTeZa94}
M. Ba{\~n}ados, C. Teitelboim, and J. Zanelli, Phys. Rev. Lett. {\bf 72},  957
  (1994), gr-qc/9309026.

\bibitem{HawHun99}
S.~W. Hawking and C.~J. Hunter, Phys. Rev. {\bf D59},  044025  (1999),
  hep-th/9808085.

\bibitem{GaGiSt94}
D. Garfinkle, S.~B. Giddings, and A. Strominger, Phys. Rev. {\bf D49},  958
  (1994), gr-qc/9306023.

\bibitem{Wald93}
R.~M. Wald, Phys. Rev. {\bf D48},  R3427  (1993), gr-qc/9307038.

\bibitem{IyeWal94}
V. Iyer and R.~M. Wald, Phys. Rev. {\bf D50},  846  (1994), gr-qc/9403028.

\bibitem{Carlip95}
S. Carlip, Phys. Rev. {\bf D51},  632  (1995), gr-qc/9409052.

\bibitem{Stromi98}
A. Strominger, JHEP {\bf 9802},  009  (1998), hep-th/9712251.

\bibitem{Carlip99}
S. Carlip, Phys. Rev. Lett. {\bf 82},  2828  (1999), hep-th/9812013.

\bibitem{BKLS86}
L. Bombelli, R.~K. Koul, J. Lee, and R.~D. Sorkin, Phys. Rev. {\bf D34},  373
  (1986).

\bibitem{Sredni93}
M. Srednicki, Phys. Rev. Lett. {\bf 71},  666  (1993), hep-th/9303048.

\bibitem{FroNov93}
V. Frolov and I. Novikov, Phys. Rev. {\bf D48},  4545  (1993), gr-qc/9309001.

\bibitem{CalWil94}
C. Callan and F. Wilczek, Phys. Lett. {\bf B333},  55  (1994), hep-th/9401072.

\bibitem{KabStr94}
D. Kabat and M.~J. Strassler, Phys. Lett. {\bf B329},  46  (1994),
  hep-th/9401125.

\bibitem{HoLaWi94}
C. Holzhey, F. Larsen, and F. Wilczek, Nucl. Phys. {\bf B424},  443  (1994),
  hep-th/9403108.

\bibitem{MuSeKo97}
S. Mukohyama, M. Seriu, and H. Kodama, Phys. Rev. {\bf D55},  7666  (1997),
  gr-qc/9701059.

\bibitem{MuSeKo98}
S. Mukohyama, M. Seriu, and H. Kodama, Phys. Rev. {\bf D58},  064001  (1998),
  gr-qc/9712018.

\bibitem{Others}
For other recent discussions on the entanglement entropy,
see e.g.,
J.~S. Dowker, Class. Quantum Grav. {\bf 11},  L55  (1994), hep-th/9401159;
F. Larsen and F. Wilczek, Ann. Phys. (N.Y.) {\bf 243},  280  (1995),
  hep-th/9408089;
R. M{\"u}ller and C.~O. Lousto, Phys. Rev. {\bf D52},  4512  (1995),
  gr-qc/9504049;
E. Benedict and S.-Y. Pi, Ann. Phys. (N.Y.) {\bf 245},  209  (1996),
  hep-th/9505121.

\bibitem{GibPer}
G.~W. Gibbons and M.~J. Perry, Phys. Rev. Lett. {\bf 36},  985  (1976);
Proc. R. Soc. London {\bf A358},  467  (1978).

\bibitem{ChrDuf78}
S.~M. Christensen and M.~J. Duff, Nucl. Phys. {\bf B146},  11  (1978).

\bibitem{TroVan79}
W. Troost and H. Van~Dam, Nucl. Phys. {\bf B152},  442  (1979).

\bibitem{Wald84}
R.~M. Wald, {\em General Relativity} (University of Chicago Press, Chicago,
  1984), p. 407.

\bibitem{GibHaw79}
G.~W. Gibbons and S.~W. Hawking, Commun. Math. Phys. {\bf 66},  291  (1979).

\bibitem{HawHor96}
S.~W. Hawking and G.~T. Horowitz, Class. Quantum Grav. {\bf 13},  1487  (1996),
  gr-qc/9501014.

\bibitem{Mukohy98}
S. Mukohyama, Phys. Rev. {\bf D58},  104023  (1998), gr-qc/9805039.

\bibitem{GraRyz80}
I.~S. Gradshteyn and I.~M. Ryzhik, {\em Table of Integrals, Series,
  and Products} (Academic Press,1980), p. 837, Eq. 7.374--8.

\bibitem{Rindle66}
W. Rindler, Am. J. Phys. {\bf 34},  1174  (1966).

\bibitem{Erdely54}
A. Erd\'elyi {\em et al.}, {\em Tables of Integral Transforms}
(McGraw-Hill, New York, 1954),
Vol. \Rn{2}, Chap. \Rn{12}.

\bibitem{Fullin73}
S.~A. Fulling, Phys. Rev. {\bf D7},  2850  (1973).

\bibitem{tHooft85}
G. 't~Hooft, Nucl. Phys. {\bf B256},  727  (1985).

\end{thebibliography}
\end{document}